\documentclass{article}
\usepackage{spconf,amsmath,graphicx}
\usepackage{hyperref}
\usepackage{cite}
\usepackage{booktabs}
\usepackage{multirow}
\usepackage{cleveref}
\usepackage{color}

\title{A Detailed Audio-Text Data Simulation Pipeline using Single-event Sounds}
\name{Xuenan Xu, Xiaohang Xu, Zeyu Xie, Pingyue Zhang, Mengyue Wu$\dag$, Kai Yu$\dag$\thanks{$\dag$ Corresponding authors. This work has been supported by National Natural Science Foundation of China (No.92048205), the Key Research and Development Program of Jiangsu Province, China (No.BE2022059), and Shanghai Municipal Science and Technology Major Project (2021SHZDZX0102). }}
\address{MoE Key Lab of Artificial Intelligence, AI Institute\\
  X-LANCE Lab, Department of Computer Science and Engineering\\
  Shanghai Jiao Tong University, Shanghai, China}

\begin{document}
%
\maketitle
\begin{abstract}
Recently, there has been an increasing focus on audio-text cross-modal learning.
However, most of the existing audio-text datasets contain only simple descriptions of sound events.
Compared with classification labels, the advantages of such descriptions are significantly limited.
In this paper, we first analyze the detailed information that human descriptions of audio may contain beyond sound event labels.
Based on the analysis, we propose an automatic pipeline for curating audio-text pairs with rich details\footnote{https://www.github.com/wsntxxn/RichDetailAudioTextSimulation}.
Leveraging the property that sounds can be mixed and concatenated in the time domain, we control details in four aspects: temporal relationship, loudness, speaker identity, and occurrence number, in simulating audio mixtures.
Corresponding details are transformed into captions by large language models.
Audio-text pairs with rich details in text descriptions are thereby obtained.
We validate the effectiveness of our pipeline with a small amount of simulated data, demonstrating that the simulated data enables models to learn detailed audio captioning.
\end{abstract}
\begin{keywords}
Detailed audio captioning, audio-text learning, data curation pipeline
\end{keywords}
%
\section{Introduction}
\label{sec:intro}

Audio-text learning has attracted increasing attention in recent years.
Similar to visual-language learning~\cite{radford2021learning,li2022blip}, machines learn audio concepts more efficiently under text supervision~\cite{elizalde2023clap}, compared with classification data where labels are pre-defined categories.
The development of audio-text cross-modal tasks~\cite{drossos2017automated,li2023diverse,oncescu2021audio,fayek2020temporal} holds promise in developing more natural human-machine interaction systems.

The limited amount of existing human-annotated data~\cite{kim2019audiocaps,drossos2020clotho,martin2021diversity} presents a significant challenge in audio-text learning.
Several works~\cite{wu2023large,xu2023blat,mei2023wavcaps} have utilized templates or generative models to convert category labels or web-crawled descriptions into captions, thereby curating larger datasets.
However, whether manually annotated or automatically curated, text descriptions are mainly limited to sound events while detailed information is neglected.
As a result, captions generated by current audio captioning models~\cite{mei2023wavcaps} only focus on sound events.
For example, models are likely to output ``a horn blares'' for the sound of two honks, neglecting details such as the occurrence time.
Previous studies~\cite{wu2023audio,huang2023make,xie2023enhance} have explored improving the correspondence of temporal relationships between audio and text. 
Besides, audio carries much more details beyond just temporal information.
Details like the occurrence number also offer unique advantages of text descriptions which cannot be acquired from category labels.

However, humans cannot always provide highly coherent, detailed captions for diverse audio samples.
Different from detailed visual information (e.g., colour and shape) in image descriptions, details in audio descriptions are ambiguous by nature.
For instance, when a sound event overlaps with others, it may be not easy to count the exact number of occurrences.
Hence such a detail-lacking phenomenon cannot be easily mitigated by only enhancing the caption quality.
In turn, the presence of incomplete or imprecise detailed audio-text pairs may adversely affect the efficacy of training.
To ensure the diversity and alignment of auditory details between audio and captions, we propose a simulation framework, where the details are controlled and captions are generated based on corresponding details.

\begin{table*}[htpb]
    \centering
    \small
    \caption{Details included in audio descriptions. Some of the sound events applicable to each specific detail are also listed, with ``all'' indicating that the detail applies to all sound events.}
    \begin{tabular}{l|l|l}
    \toprule
    \textbf{Detail}  & \textbf{Applicable Sound Categories} & \textbf{Caption Example}\\
    \midrule
    Occurrence number & Gunshot, dog barking (short events) & A gun is fired \textbf{\textit{three times}}. \\
    Identity of sounding objects & Man speaking, woman laughing & A man speaks while \textbf{\textit{another man}} responds. \\
    Duration & water running, spraying & Water running \textbf{\textit{continuously}}. \\
    Temporal relationship & All & A man speaks \textbf{\textit{then}} coughs \textbf{\textit{followed by}} laughter. \\
    Loudness & All & Bells ringing \textit{\textbf{loudly}}.\\
    Frequency / Pitch & Engine humming, beeping & A \textbf{\textit{high pitched}} engine running consistently\\
    Velocity & Gunshot, keyboard typing & Someone types on a keyboard \textit{\textbf{super fast}}\\
    Audio quality & Radio & Person chatting with \textit{\textbf{poor audio record}}\\
    Environment & - & A horse neighs and birds chirp \textit{\textbf{outside}}\\
    Emotion & Insect buzzing, bird chirping & A very loud and \textbf{\textit{annoying}} fire truck siren\\
    \bottomrule
    \end{tabular}
    \label{tab:details_summarization}
\end{table*}

An auditory detail taxonomy is built based on AudioCaps~\cite{kim2019audiocaps}, which covers common non-music natural sound events, to control details in audio-text simulation.
First, we establish a detailed sound event taxonomy, encompassing the auditory details that can be described, along with the corresponding sound categories since some details are bound with certain categories.
Such an auditory detail taxonomy is achieved by analyzing human annotations and clustering on AudioCaps corpus.
Then, by analyzing AudioCaps captions, we summarize common sound description details and applicable sound events of each detail.

Based on the taxonomy, we propose an automatic pipeline for generating detailed audio-text data.
Single-event sounds are first curated from the Freesound platform~\cite{font2013freesound}. 
Multi-event audio is simulated by mixing single events, similar to approaches in speech processing~\cite{landini2022from}.
The metadata of the simulation process, like occurrence numbers, is converted into natural text by ChatGPT.

In this way, detailed and fully-aligned audio-text data is curated.
As a proof-of-concept, we validate the effect of simulated data on audio captioning.
By fine-tuning on a small amount of simulated data, the captioning model is capable of generating more detailed captions.
Our proposed pipeline is not limited to enhancing captioning but also holds promise for improving audio-text correspondence in terms of details in audio generation and cross-modal understanding. 


\section{Auditory Detail Taxonomy based on Human Perception}
\label{sec:category_detail}

To generate an audio-text pair with rich details, taxonomies of sound event categories and details are required.
We construct both systems from human-annotated audio captioning dataset, AudioCaps, for consistency with human perception.

\subsection{Sound Event Categories from Text Clustering}

Although existing classification datasets provide sound event category systems~\cite{gemmeke2017audio,chen2020vggsound}, there are two main issues in these categories: 1) There are overlaps between different categories (e.g., ``Liquid'' and ``Water'').
2) Some acoustically similar categories are treated as different events, even though they are indistinguishable to human (e.g., ``Crackle'' and ``Crumpling'').

Due to these two issues, simulating audio-text data based on existing sound categories may lead to text descriptions with information unavailable from the audio.
This poses challenges for training audio captioning models, as the model is trained to generate disparate captions for similar sounds.
To avoid curating such descriptions, we do clustering on human-annotated captions to summarize categories based on human perception.

We first extract sound event phrases by splitting AudioCaps captions using pre-defined delimiters~\cite{xu2023investigating}.
Then, phrases are transformed into CLAP~\cite{wu2023large} embeddings so that descriptions of acoustically similar events are close in the embedding space.
K-means clustering is applied to group descriptions of similar sound events.
We use an empirical cluster number of 64 so that categories are neither too fine-grained nor coarse-grained.
A representative label is assigned to each cluster by querying ChatGPT with a few phrases.

\subsection{Details in Sound Descriptions}

To summarize sound details, part-of-speech (POS) tagging is applied to AudioCaps captions to extract adjectives and adverbs.
These words are then summarized into several description aspects, shown in \Cref{tab:details_summarization}.
Both attributes of single events (e.g., duration, loudness) and relationships involving different events (e.g. temporal relationships) are included.
Some details are not listed since they are only applicable to speech or music, e.g., the speech content and language.
These details are difficult to recognize without specific models.

\begin{figure*}
    \centering
    \includegraphics[width=0.95\textwidth]{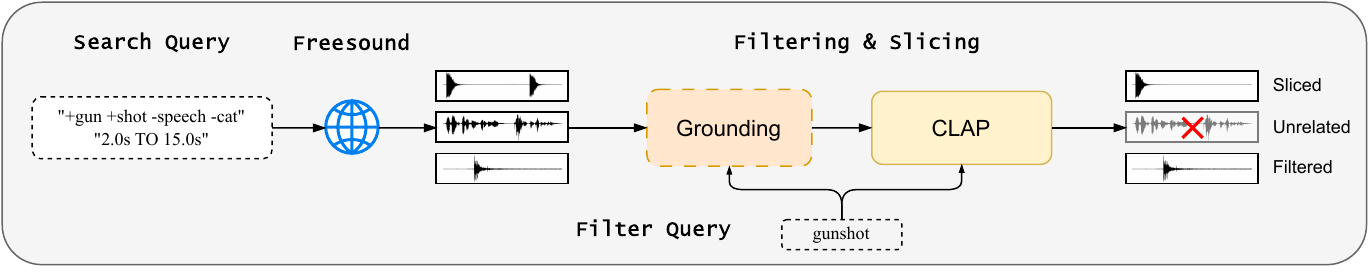}
    \caption{The pipeline of single-event sound collection from Freesound.}
    \label{fig:freesound_curation}
\end{figure*}

\begin{figure*}
    \centering
    \includegraphics[width=0.95\textwidth]{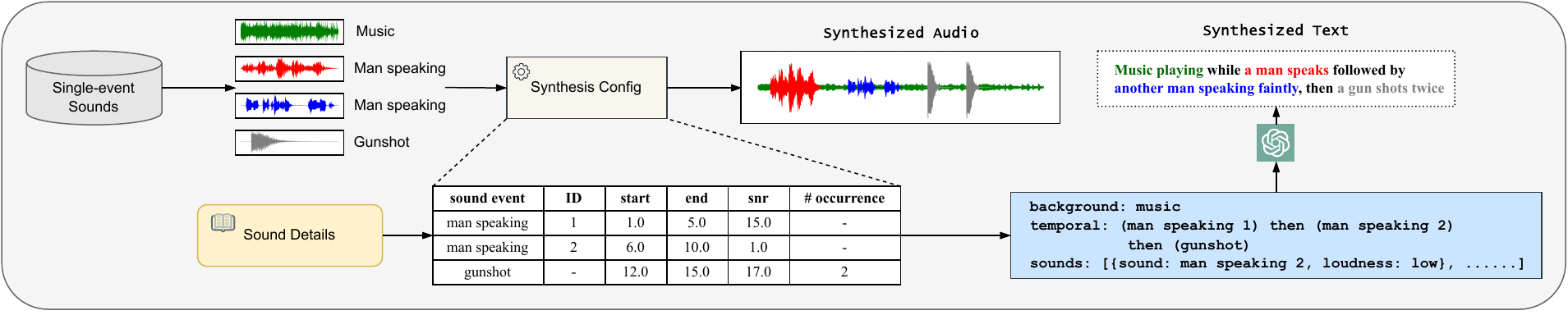}
    \caption{The pipeline of simulating audio-text pairs that are rich in details using automatically-curated single sound sources and large language models.}
    \label{fig:synthesis_pipeline}
\end{figure*}

It should be noted that some details are specific to sound event categories.
For example, ``occurrence number'' is only applicable to short-duration events like gunshots or dog barking.
``Identity'' is only discernable for events like man speaking while humming from different machines of the same model are often indistinguishable for humans.
Therefore, we also summarize details applicable to each sound event for detail controlling in simulation.

\section{Detailed Audio-Text Simulation Pipeline}
\label{sec:synthesis_pipeline}

\subsection{Single-event Sound Curation}

Although data of some sound event categories are available in classification datasets, target sound events are often accompanied by background noise.
In contrast, single-event sounds on Freesound are often clean.
Therefore, we collect sound event data from Freesound.
As \Cref{fig:freesound_curation} shows, we write search queries for each sound event to download relevant audio clips.
Queries are based on clustering labels.
Clips that are too short or too long are discarded.
Then, two models are adopted for filtering.
A TAG model~\cite{xu2023investigating} detects when the target sound event occurs.
It filters out clips that contain too long non-target segments or splits them into several segments.
Then, CLAP calculates the similarity score between an audio clip and the event description so audio clips with low scores are filtered out.
Both thresholds are emperically selected based on a few examples.
For short-duration events, only CLAP filtering is applied.
We take a tradeoff between data quality and quantity using such an automatic pipeline.
Finally, we curate audio clips for a total of 115 sound events and most events contain more than 50 samples.

\subsection{Audio-Text Simulation Pipeline}
Based on curated single events and summarized audio details, we propose a detailed audio-text data simulation pipeline.
As presented in \Cref{fig:synthesis_pipeline}, we randomly sample sound events and corresponding samples from the curated pool.
According to details that sampled events are related to, we randomly sample attributes.
Then, the audio mixture is simulated based on the configuration and single events.
With an empirical probability, we do not add background noise to simulate situations where no discernible source is continuously present in the real dataset.
The configuration is also transformed into a structured metadata denoting details.
Finally, ChatGPT transforms the metadata into a descriptive caption with the style of human annotations.
In this way, fully-aligned audio-text pairs with rich details are obtained by simulation.

\section{Experimental Setup}
\label{sec:exp_setup}

\begin{table*}[htpb]
    \centering
    \small
    \caption{Examples of generated captions from the baseline model and fine-tuned model.}
    \begin{tabular}{c|c|c}
    \toprule
    Audio & YDn3buZWMzwY.wav & Y2j8pxiFvElM.wav \\
    \midrule
     & \includegraphics[width=0.35\textwidth]{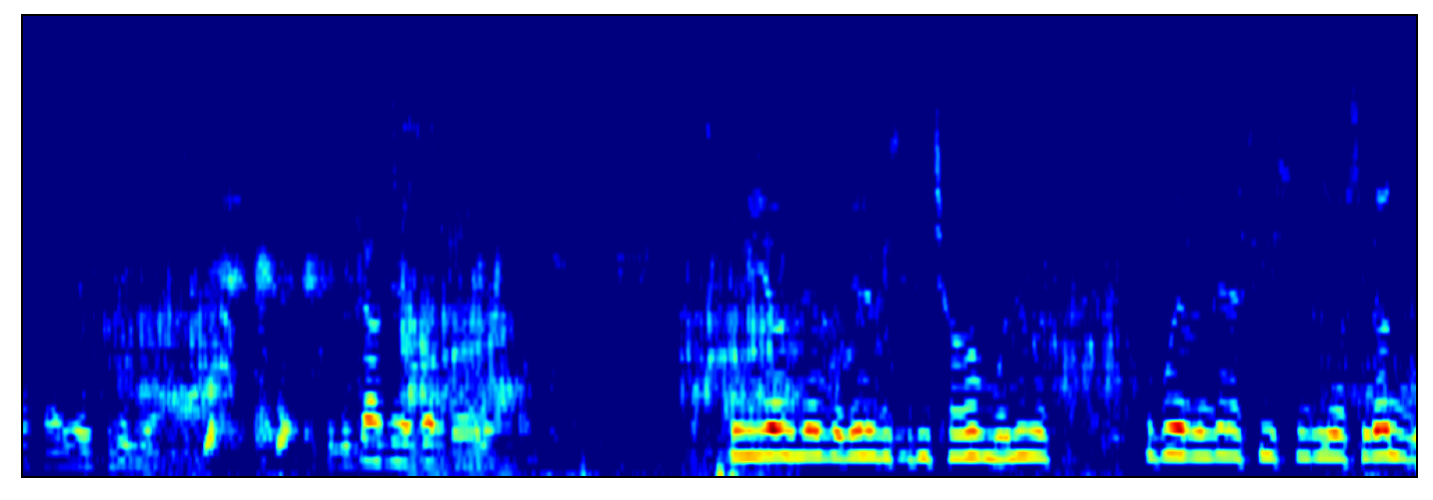} & \includegraphics[width=0.35\textwidth]{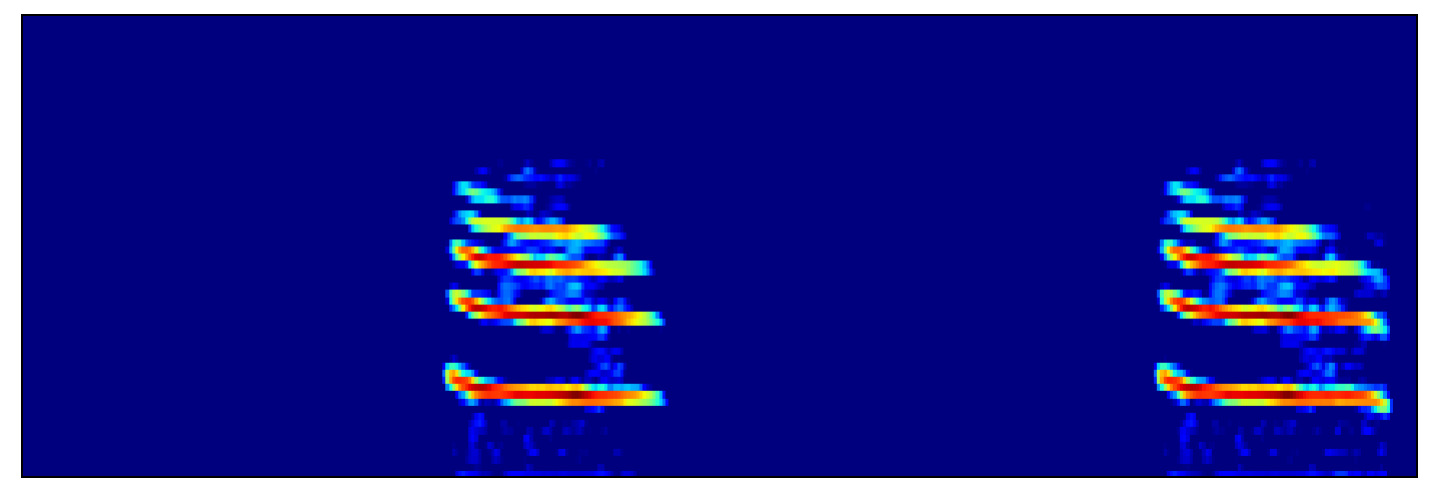} \\
    \hline
    Baseline & a man talks while a sewing machine works & a cat meows \\
    \hline
    \multirow{2}{*}{Fine-tuned} & a man speaks \texttt{followed by another} & \multirow{2}{*}{a cat meows \texttt{twice}} \\
     & man speaking and a woman laughing & \\
    \hline
    \multirow{3}{*}{References} & men speak as someone snores & a cat meowing twice \\
    \cline{2-3}
    & snoring is ongoing, while an adult female & \multirow{2}{*}{a cat meows repeatly}\\
    & laughs and an adult male speaks & \\
    \bottomrule
    \end{tabular}
    \label{tab:examples}
\end{table*}

\subsection{Data Simulation}
Since the quantity is not large enough to support large-scale simulation due to the filtering, we simulate a small amount of data to prevent using the same sound sources many times, which will lead to over-fitting.
As a proof-of-concept, 4 aspects of details are controlled during simulation: occurrence number, identity, loudness and temporal relationships.
For less straightforward detail like emotion, we would need a sound-induced emotion classification model to provide the clue, which we leave in the future work.
We simulate two datasets.
All sound event categories are involved to simulate the first one.
Categories with over 200 curated sources are used in the second to allow more simulated samples.
To improve the quality and detailedness of simulated data, we filter out audio-text pairs without keywords related to details (e.g., ``twice'' and ``another'') in the text and with low CLAP scores.
Finally, the combination of two datasets yield 2,785 simulated audio-text pairs for fine-tuning.

\subsection{Hyper-parameters}

Since the amount of simulated data is very small, we fine-tune a captioning model pre-trained on AudioCaps instead of training from scratch.
We take the same architecture as \cite{xu2022sjtu}.
The model trained on AudioCaps without fine-tuning is taken as the baseline.
Due to the small dataset size, we use a very small learning rate of $2\times10^{-6}$ and apply augmentations like random noise and pitch shift on the original audio to alleviate over-fitting.
The model is trained for 20 epochs.

\section{Results}
\label{sec:results}

\subsection{Objective Metrics}
\label{subsec:obj_metric}
First, we evaluate the effectiveness of simulated data using objective metrics.
Captioning metrics including BLEU-4 ($\text{B}_\text{4}$), ROUGE (R), METEOR (M), SPIDEr (SD) and SentenceBERT (SB) are calculated.
Compared with FENSE, we do not employ grammar error penalties to focus on only description accuracy.
The comparison shows a consistent slight improvement of the fine-tuned model over the baseline.
Additionally, we adopt temporal $\text{F}_\text{1}$~\cite{xie2023enhance} to evaluate the accuracy of temporal relationships.
Significant improvement is achieved by fine-tuning on small-scale simulated data with rich temporal details.
However, for other details like identity and occurrence number, there are no appropriate metrics so we conduct human evaluation.
\begin{table}[htpb]
    \centering
    \small
    \caption{Performance comparison using objective metrics.}
    \begin{tabular}{c|ccccccc}
    \toprule
    Model & $\text{B}_\text{4}$ & R & M & SD & SB & $\text{F}_\text{1-temp}$\\
    \midrule
    Baseline & 26.9 & 49.7 & 24.5 & 46.2 & 62.3 & 35.1\\
    Fine-tuned & \textbf{27.7} & \textbf{51.0} & \textbf{26.1} & \textbf{47.1} & \textbf{62.6} & \textbf{60.9}\\
    \bottomrule
    \end{tabular}
    \label{tab:objective_metrics}
\end{table}

\subsection{Human Evaluation}
We randomly choose 20 samples whose captions generated by the baseline and fine-tuned model are disparate and invite 10 raters to choose which one is more accurate and more detailed, respectively.
The result shows that captions generated by the fine-tuned model are likely to be more accurate than those generated by the baseline model.
Moreover, in most cases, the fine-tuned model generates captions that contain more details than the baseline model generated.
We provide two examples in \Cref{tab:examples}.
Sound events in captions generated by both models are often similar.
However, the fine-tuned model's predictions include more details, such as identity (``another man'') and occurrence number (``twice''), making them more accurate.
We also observe that due to the distribution difference in simulated data and AudioCaps, the fine-tuned model may also generate details that are not available from the input audio.
We will work on this problem in the future work.
\begin{table}[htpb]
    \centering
    \small
    \caption{Win rate of different models in human evaluation.}
    \begin{tabular}{c|cc}
    \toprule
    Model & Accuracy & Detailedness\\
    \midrule
    Baseline & 30.0 & 9.5 \\
    Fine-tuned & 62.0 & 82.0 \\
    \bottomrule
    \end{tabular}
    \label{tab:human_eval}
\end{table}

\vspace{-2mm}

\section{Conclusion}
\label{sec:conclusion}

In this work, we try to tackle a significant problem in audio-text learning that details are lacking in existing data.
We summarize sound event categories and details by analyzing text descriptions in AudioCaps.
Based on the clustered event categories and corresponding details, we propose an automatic pipeline to simulate audio-text pairs with rich details.
We adopt a TAG model and CLAP to filter high-quality single events from Freesound.
Then, sound mixtures are simulated from curated sounds with specific controls on details.
The associated metadata is fed to ChatGPT to generate natural language descriptions of the simulated audio with details.
By simulating a small amount of data controlling 4 aspects of details, we conduct a proof-of-concept.
Experimental results show that the model fine-tuned on simulated data is able to generate more detailed captions.

\bibliographystyle{IEEEtran}
\bibliography{refs}

\begin{thebibliography}{10}
\providecommand{\url}[1]{#1}
\csname url@samestyle\endcsname
\providecommand{\newblock}{\relax}
\providecommand{\bibinfo}[2]{#2}
\providecommand{\BIBentrySTDinterwordspacing}{\spaceskip=0pt\relax}
\providecommand{\BIBentryALTinterwordstretchfactor}{4}
\providecommand{\BIBentryALTinterwordspacing}{\spaceskip=\fontdimen2\font plus
\BIBentryALTinterwordstretchfactor\fontdimen3\font minus \fontdimen4\font\relax}
\providecommand{\BIBforeignlanguage}[2]{{%
\expandafter\ifx\csname l@#1\endcsname\relax
\typeout{** WARNING: IEEEtran.bst: No hyphenation pattern has been}%
\typeout{** loaded for the language `#1'. Using the pattern for}%
\typeout{** the default language instead.}%
\else
\language=\csname l@#1\endcsname
\fi
#2}}
\providecommand{\BIBdecl}{\relax}
\BIBdecl

\bibitem{radford2021learning}
A.~Radford, J.~W. Kim, C.~Hallacy, A.~Ramesh, G.~Goh, S.~Agarwal, G.~Sastry, A.~Askell, P.~Mishkin, J.~Clark \emph{et~al.}, ``Learning transferable visual models from natural language supervision,'' in \emph{Proc. ICML}, 2021, pp. 8748--8763.

\bibitem{li2022blip}
J.~Li, D.~Li, C.~Xiong, and S.~Hoi, ``Blip: Bootstrapping language-image pre-training for unified vision-language understanding and generation,'' in \emph{International Conference on Machine Learning}, 2022, pp. 12\,888--12\,900.

\bibitem{elizalde2023clap}
B.~Elizalde, S.~Deshmukh, M.~Al~Ismail, and H.~Wang, ``Clap learning audio concepts from natural language supervision,'' in \emph{Proc. IEEE ICASSP}, 2023, pp. 1--5.

\bibitem{drossos2017automated}
K.~Drossos, S.~Adavanne, and T.~Virtanen, ``Automated audio captioning with recurrent neural networks,'' in \emph{Proc. IEEE WASPAA}, 2017, pp. 374--378.

\bibitem{li2023diverse}
G.~Li, X.~Xu, L.~Dai, M.~Wu, and K.~Yu, ``Diverse and vivid sound generation from text descriptions,'' in \emph{Proc. IEEE ICASSP}, 2023, pp. 1--5.

\bibitem{oncescu2021audio}
A.-M. Oncescu, A.~Koepke, J.~F. Henriques, Z.~Akata, and S.~Albanie, ``Audio retrieval with natural language queries,'' in \emph{Proc. ISCA Interspeech}, 2021, pp. 2411--2415.

\bibitem{fayek2020temporal}
H.~M. Fayek and J.~Johnson, ``Temporal reasoning via audio question answering,'' \emph{IEEE/ACM TASLP}, vol.~28, pp. 2283--2294, 2020.

\bibitem{kim2019audiocaps}
C.~D. Kim, B.~Kim, H.~Lee, and G.~Kim, ``Audiocaps: Generating captions for audios in the wild,'' in \emph{Proc. NAACL}, 2019, pp. 119--132.

\bibitem{drossos2020clotho}
K.~Drossos, S.~Lipping, and T.~Virtanen, ``Clotho: An audio captioning dataset,'' in \emph{Proc. IEEE ICASSP}, 2020, pp. 736--740.

\bibitem{martin2021diversity}
I.~Martin and A.~Mesaros, ``Diversity and bias in audio captioning datasets,'' in \emph{Proc. DCASE}, 2021, pp. 90--94.

\bibitem{wu2023large}
Y.~Wu, K.~Chen, T.~Zhang, Y.~Hui, T.~Berg-Kirkpatrick, and S.~Dubnov, ``Large-scale contrastive language-audio pretraining with feature fusion and keyword-to-caption augmentation,'' in \emph{Proc. IEEE ICASSP}, 2023, pp. 1--5.

\bibitem{xu2023blat}
X.~Xu, Z.~Zhang, Z.~Zhou, P.~Zhang, Z.~Xie, M.~Wu, and K.~Q. Zhu, ``Blat: Bootstrapping language-audio pre-training based on audioset tag-guided synthetic data,'' \emph{arXiv preprint arXiv:2303.07902}, 2023.

\bibitem{mei2023wavcaps}
X.~Mei, C.~Meng, H.~Liu, Q.~Kong, T.~Ko, C.~Zhao, M.~D. Plumbley, Y.~Zou, and W.~Wang, ``Wavcaps: A chatgpt-assisted weakly-labelled audio captioning dataset for audio-language multimodal research,'' \emph{arXiv preprint arXiv:2303.17395}, 2023.

\bibitem{wu2023audio}
H.-H. Wu, O.~Nieto, J.~P. Bello, and J.~Salomon, ``Audio-text models do not yet leverage natural language,'' in \emph{Proc. IEEE ICASSP}.\hskip 1em plus 0.5em minus 0.4em\relax IEEE, 2023, pp. 1--5.

\bibitem{huang2023make}
J.~Huang, Y.~Ren, R.~Huang, D.~Yang, Z.~Ye, C.~Zhang, J.~Liu, X.~Yin, Z.~Ma, and Z.~Zhao, ``Make-an-audio 2: Temporal-enhanced text-to-audio generation,'' \emph{arXiv preprint arXiv:2305.18474}, 2023.

\bibitem{xie2023enhance}
Z.~Xie, X.~Xu, M.~Wu, and K.~Yu, ``Enhance temporal relations in audio captioning with sound event detection,'' in \emph{Proc. ISCA Interspeech}, 2023, pp. 4179--4183.

\bibitem{font2013freesound}
F.~Font, G.~Roma, and X.~Serra, ``Freesound technical demo,'' in \emph{Proc. ACM MM}, 2013, pp. 411--412.

\bibitem{landini2022from}
F.~Landini, A.~Lozano-Diez, M.~Diez, and L.~Burget, ``From simulated mixtures to simulated conversations as training data for end-to-end neural diarization,'' in \emph{Proc. ISCA Interspeech}, 2022, pp. 5095--5099.

\bibitem{gemmeke2017audio}
J.~F. Gemmeke, D.~P. Ellis, D.~Freedman, A.~Jansen, W.~Lawrence, R.~C. Moore, M.~Plakal, and M.~Ritter, ``Audio set: An ontology and human-labeled dataset for audio events,'' in \emph{Proc. IEEE ICASSP}, 2017, pp. 776--780.

\bibitem{chen2020vggsound}
H.~Chen, W.~Xie, A.~Vedaldi, and A.~Zisserman, ``Vggsound: A large-scale audio-visual dataset,'' in \emph{Proc. IEEE ICASSP}, 2020, pp. 721--725.

\bibitem{xu2023investigating}
X.~Xu, M.~Wu, and K.~Yu, ``Investigating pooling strategies and loss functions for weakly-supervised text-to-audio grounding via contrastive learning,'' in \emph{Proc. IEEE ICASSPW}, 2023, pp. 1--5.

\bibitem{xu2022sjtu}
X.~Xu, Z.~Xie, M.~Wu, and K.~Yu, ``The {SJTU} system for {DCASE2022} challenge task 6: Audio captioning with audio-text retrieval pre-training,'' DCASE2022 Challenge, Tech. Rep., 2022.

\end{thebibliography}

\end{document}